\documentclass[pdflatex,sn-mathphys-num]{sn-jnl}

\usepackage{graphicx}%
\usepackage{multirow}%
\usepackage{amsmath,amssymb,amsfonts}%
\usepackage{amsthm}%
\usepackage[cal=rsfs]{mathalfa}
\usepackage[title]{appendix}%
\usepackage{xcolor}%
\usepackage{textcomp}%
\usepackage{manyfoot}%
\usepackage{booktabs}%
\usepackage{algorithm}%
\usepackage{algorithmicx}%
\usepackage{algpseudocode}%
\usepackage{listings}%
\usepackage{float}
\usepackage{tikz}
\usetikzlibrary{decorations.pathreplacing}
\usepackage{forest}
\usepackage{caption}
\usetikzlibrary{arrows.meta, positioning}

\usepackage{comment}

\theoremstyle{thmstyleone}%
\theoremstyle{thmstyletwo}%

\theoremstyle{thmstylethree}%
\newtheorem{definition}{Definition}%

\begin{document}

\title[MergeLLL]{MergeLLL: A Hierarchical Divide-and-Conquer Framework for LLL-Based Lattice Reduction}


\author*[1]{\fnm{Niharika} \sur{Gauraha}}\email{niharika@kth.se}



\affil*[1]{\orgdiv{Department of Theoretical Computer Science}, \orgname{KTH, The Royal Institute of Technology}, \country{Stockholm}}




\abstract{
Lattice basis reduction algorithms have various applications in computational number theory and lattice-based cryptography, but their complexity increases rapidly with the dimension. Motivated by the divide-and-conquer strategy of merge sort and incorporating PotLLL-style deep insertions during recombination, MergeLLL is proposed. In this framework, a lattice basis is split into sub-bases (blocks), where local reductions are performed within each block using KZ reduction, approximated in practice by BKZ with block size equal to the sublattice dimension. The full basis is then reconstructed through a hierarchical merging process.

The approach focuses on improving local lattice structure before refining global basis properties. The method is naturally parallelizable, enabling efficient multicore and distributed execution. The reduction and merging steps preserve the lattice structure via unimodular transformations and admit logarithmic parallel depth. In experiments on subset-sum and NTRU-style lattices, the method is evaluated against PotLLL and BKZ 2.0, showing comparable computation time with PotLLL while achieving a lower root Hermite factor, indicating higher-quality reduced bases.

}




\maketitle

\section{Introduction} 
Lattice basis reduction is a fundamental tool in several fields, including number theory and modern cryptography. Given a lattice basis, reduction algorithms compute an alternative basis with shorter and more nearly orthogonal vectors while preserving the underlying lattice structure. Since the introduction of the Lenstra–Lenstra–Lovász (LLL) algorithm \cite{lll1982}, lattice reduction has become a central component of cryptanalysis, integer programming, Diophantine approximation, and, more recently, the analysis and implementation of post-quantum cryptographic systems \cite{micciancio2002complexity,NguyenVallee2010}.
This growing range of applications has also motivated the development of more efficient and scalable reduction algorithms tailored to high-dimensional settings.

Among existing reduction techniques, block Korkine–Zolotarev (BKZ) \cite{SchnorrEuchner1994,BKZ2011} reduction improves over LLL by operating on blocks of the basis, while PotLLL\cite{PotLLL2014} enhances LLL through deep insertions guided by potential minimization, enabling higher-quality reductions in practice.

Most lattice reduction algorithms operate globally on the full basis, capturing dependencies among all vectors but at high computational cost. To reduce the computational overhead associated with maintaining Gram–Schmidt orthogonalization (GSO) data during lattice reduction, several authors have proposed segment- and block-based approaches\cite{schnorr2006fast, schnorr1994blockwise} . In Segment LLL, the basis is partitioned into overlapping segments that are reduced largely independently before being reconciled recursively. While modern lattice reduction libraries\cite{NguyenVallee2010} often employ incremental and lazy-update strategies rather than explicit Segment LLL, the underlying principle remains the same: restricting orthogonalization work to local blocks can significantly improve practical performance without substantially affecting reduction quality.

In this work, MergeLLL is introduced as a hierarchical divide-and-conquer framework for lattice basis reduction. Inspired by the decomposition and recombination structure of merge sort, the full basis is not treated as a single object but instead partitioned into smaller sub-bases. Independent local reductions are performed within each sub-basis using KZ reduction, approximated by BKZ with block size equal to the sub-basis dimension, and the resulting components are merged through a hierarchical reduction tree.

The central idea behind MergeLLL is that local lattice structure can often be improved before global interactions are considered. Local reductions shorten vectors, improve orthogonality, and reduce conditioning issues within lower-dimensional subspaces. By performing these operations independently, much of the computation can be parallelized. Global structure is then reintroduced through a sequence of merge operations, allowing information to propagate across partitions while preserving the gains from local reduction.

A key contribution of this work is a novel hierarchical merging procedure for lattice bases. While the overall framework follows the divide-and-conquer structure of merge sort, the merge step itself constitutes the main algorithmic innovation. Inspired by PotLLL reduction \cite{PotLLL2014}, the proposed merge strategy goes beyond simply concatenating independently reduced sub-bases. Instead, it performs structured insertion and reordering operations that progressively restore interactions between partitions and improve basis quality at each level of the hierarchy. 

From a theoretical perspective, MergeLLL preserves the underlying lattice throughout the reduction process. Each local reduction and merge step consists of unimodular basis transformations, ensuring that every intermediate and final basis generates the same lattice as the original input. Moreover, the hierarchical reduction tree exposes substantial parallelism: for a balanced decomposition, the merge hierarchy has logarithmic depth, while most reduction work is performed independently on disjoint sub-bases.

To evaluate the effectiveness of the proposed framework, both classical and cryptographically relevant lattice families are considered. In particular, lattices arising from subset-sum instances \cite{SchnorrEuchner1994} and structured lattices derived from NTRU-style constructions \cite{hoffstein1998ntru} are studied. These classes are characterized by distinct reduction challenges and provide insight into the behavior of the hierarchical method on both unstructured and highly structured lattices. The resulting bases are assessed using the root Hermite factor (RHF). The results show that basis quality is improved by MergeLLL and scalable parallel execution in high-dimensional settings is naturally enabled by its divide-and-conquer structure.

The main contributions of this paper are summarized as follows:
\begin{itemize}
  \item MergeLLL is introduced as a divide-and-conquer framework for lattice basis reduction inspired by the merge-sort paradigm. Each sub-basis is reduced using block-wise BKZ as an approximation of KZ reduction, and a hierarchical merging strategy is proposed, incorporating PotLLL-style deep insertions to progressively refine interactions between independently reduced sub-bases.
    
    \item Theoretical properties of the framework are established, including lattice preservation under unimodular transformations and logarithmic parallel depth for balanced decompositions.
    
    \item The proposed approach is experimentally evaluated on subset-sum and NTRU-style lattices, demonstrating improved root Hermite factor.
    
    \item It is shown that the hierarchical approach provides an effective and scalable method for lattice reduction in high-dimensional settings, particularly in parallel computing environments.
\end{itemize}

\section{Preliminaries} \label{sec:preliminaries}

In this section, the definitions and results from lattice theory and lattice basis reduction that are used throughout the paper are briefly recalled.

\subsection{Lattices and Bases}

Let $\mathbf{b}_1,\mathbf{b}_2,\ldots,\mathbf{b}_n$ be vectors in $\mathbb{R}^n$. Consider the set of all integer linear combinations of these vectors:
\[
L=\left\{\sum_{i=1}^{n} a_i\mathbf{b}_i \;\middle|\; a_i\in\mathbb{Z}\right\}.
\]

Our goal is to study the structure of $L$ and, in particular, to find vectors in $L$ whose Euclidean length is small.

\begin{definition}[Lattice]
A \emph{lattice} in $\mathbb{R}^n$ is the set of all integer linear combinations of a basis
$\mathbf{b}_1,\mathbf{b}_2,\ldots,\mathbf{b}_n$ of $\mathbb{R}^n$, namely
\[
L=\left\{\sum_{i=1}^{n} a_i\mathbf{b}_i \;\middle|\; a_i\in\mathbb{Z}\right\}.
\]
Equivalently, a lattice is a discrete subgroup of $\mathbb{R}^n$ generated by a basis of $\mathbb{R}^n$.
\end{definition}

Throughout this discussion, it is assumed for simplicity that
$\mathbf{b}_1,\mathbf{b}_2,\ldots,\mathbf{b}_n \in \mathbb{Z}^n$
are linearly independent. Since $n$ linearly independent vectors in
$\mathbb{R}^n$ form a basis of $\mathbb{R}^n$, these vectors generate a
full-rank lattice $L \subseteq \mathbb{R}^n$.

Let $L$ be a lattice with basis
$\mathbf{b}_1,\mathbf{b}_2,\ldots,\mathbf{b}_n$. Associated with this basis is the
\emph{basis matrix}
\[
B=
\begin{pmatrix}
\mathbf{b}_1\\
\mathbf{b}_2\\
\vdots\\
\mathbf{b}_n
\end{pmatrix},
\]
whose rows are the basis vectors of the lattice.

Let $L \subset \mathbb{R}^n$ be a full-rank lattice with basis $b_1, \ldots, b_n$, and let $B$ be its basis matrix. If $B'$ is another basis of $L$, then there exists a unimodular matrix $U \in \mathbb{Z}^{n \times n}$ such that
\[
B' = U B,
\]
where $U$ is unimodular, i.e., $U$ is an integer matrix with $\det(U) = \pm 1$.

Since $|\det(U)|=1$, it follows that
\[
|\det(B')|
=
|\det(U)\det(B)|
=
|\det(B)|.
\]
Hence the quantity $|\det(B)|$ is independent of the choice of basis.

\subsection{Gram--Schmidt Orthogonalization}

    Given a basis $(\mathbf{b}_1, \dots, \mathbf{b}_n)$, the Gram--Schmidt process produces an orthogonal family $(\mathbf{b}_1^*, \dots, \mathbf{b}_n^*)$ defined recursively by
    \[
    \mathbf{b}_i^* = \mathbf{b}_i - \sum_{j < i} \mu_{i,j}\mathbf{b}_j^*,
    \]
    where
    \[
    \mu_{i,j}
    =
    \frac{\langle \mathbf{b}_i,\mathbf{b}_j^*\rangle}
    {\|\mathbf{b}_j^*\|^2}
    \]
    are the \emph{Gram--Schmidt coefficients}. The vectors
    $\mathbf{b}_1^*, \dots, \mathbf{b}_n^*$ are mutually orthogonal. 

\subsection{The LLL Algorithm}

Let $B=(\mathbf{b}_1,\ldots,\mathbf{b}_n)$ be a lattice basis and let
$(\mathbf{b}_1^*,\ldots,\mathbf{b}_n^*)$ denote its Gram--Schmidt
orthogonalization. 
The Lenstra--Lenstra--Lovász (LLL) algorithm produces a reduced basis satisfying size reduction and the Lovász condition. For a parameter $\delta \in (1/4,1)$, a basis is LLL-reduced if for all $i$,
\[
|\mu_{i,j}| \le \frac{1}{2}, \quad j < i,
\]
and
\[
\delta \lVert \mathbf{b}_{i-1}^* \rVert^2 \le \lVert \mathbf{b}_i^* \rVert^2 + \mu_{i,i-1}^2 \lVert \mathbf{b}_{i-1}^* \rVert^2.
\]
For details we refer to the original paper \cite{lll1982}.    

\subsection{KZ and BKZ Lattice Reduction}
Korkine--Zolotarev (KZ), Block KZ (BKZ), and BKZ 2.0 reduction are briefly recalled; for details, reference is made to the original works~\cite{SchnorrEuchner1994, BKZ2011}. In this work, BKZ 2.0 is used as a black-box approximation of BKZ, and we do not discuss its internal structure further.

\subsubsection*{KZ reduction:\\}
Korkine--Zolotarev (KZ) reduction constructs a basis iteratively by solving exact SVP instances in projected lattices. At step $k$, letting $\pi_k$ denote the projection orthogonal to $\mathrm{span}(b_1^*,\dots,b_{k-1}^*)$, the vector $b_k$ is chosen as
\[
b_k \in \arg\min_{v \in \pi_k({L}) \setminus \{0\}} \|v\|,
\]
and inserted into the basis, followed by size reduction. KZ is optimal but computationally infeasible due to repeated exact SVP calls.

\subsubsection*{BKZ reduction:\\}
Block Korkine--Zolotarev (BKZ) reduction relaxes KZ by restricting SVP computations to blocks of size $\beta$. For each $k$, an approximate shortest vector is computed in the projected block lattice generated by $(b_k,\dots,b_{k+\beta-1})$. If a shorter vector than $b_k^*$ is found, it is inserted and the basis is size-reduced. Repeating over all $k$ defines a BKZ tour. BKZ interpolates between LLL ($\beta=2$) and exact SVP/KZ ($\beta=n$), trading off quality and runtime.


\subsubsection*{BKZ 2.0 and fast implementation:\\}
BKZ 2.0 is a practical refinement of BKZ that improves the efficiency of the block SVP oracle through pruned enumeration and heuristic cost models. It preserves the BKZ framework, which interpolates between LLL and KZ-like reduction, while significantly reducing the computational cost of achieving comparable reduction quality in practice.

\subsection{The PotLLL Algorithm}

The PotLLL algorithm is a variant of lattice basis reduction based on a
global potential function rather than purely local conditions on the
Gram--Schmidt coefficients. It was introduced to improve the practical
behavior of LLL-type reductions by allowing more flexible swaps while
maintaining polynomial-time termination.

\begin{algorithm}[H]
\caption{Potential LLL (PotLLL), detailed version} \label{algo:potLLL}
\begin{algorithmic}[1]
\Require Basis $B \in \mathbb{Z}^{n\times m}$, $\delta \in (1/4,1]$
\Ensure A $\delta$-PotLLL reduced basis

\State $\delta$-LLL reduce $B$
\State $\ell \gets 1$

\While{$\ell \le n$}
    \State Size-reduce $b_\ell$ by $b_1,\ldots,b_{\ell-1}$
    \State Update $\|b_\ell^*\|^2$ and $\mu_{\ell,j}$ for $1 \le j < \ell$

    \State $P \gets 1$
    \State $P_{\min} \gets 1$
    \State $k \gets 1$

    \For{$j=\ell-1$ downto $1$}
        \State
        $P \gets P \cdot
        \dfrac{
            \|b_\ell^*\|^2 +
            \sum_{i=j}^{\ell-1}
            \mu_{\ell,i}^2 \|b_i^*\|^2
        }{
            \|b_j^*\|^2
        }$
        \If{$P < P_{\min}$}
            \State $k \gets j$
            \State $P_{\min} \gets P$
        \EndIf
    \EndFor

    \If{$\delta > P_{\min}$}
        \State $B \gets \sigma_{k,\ell} B$
        \State Update $\|b_k^*\|^2$ and $\mu_{k,j}$ for $1 \le j < k$
        \State $\ell \gets k$
    \Else
        \State $\ell \gets \ell + 1$
    \EndIf
\EndWhile

\State \Return $B$
\end{algorithmic}
\end{algorithm}

Let $B=(\mathbf{b}_1,\ldots,\mathbf{b}_n)$ be a lattice basis with
Gram--Schmidt vectors $(\mathbf{b}_1^*,\ldots,\mathbf{b}_n^*)$. The
\emph{potential} of $B$ is defined by
\[
\operatorname{Pot}(B)
=
\prod_{i=1}^{n}
\|\mathbf{b}_i^*\|^{2(n-i+1)}.
\]

The central idea of PotLLL is to perform size reduction as in LLL, followed by allowing more general permutations of basis vectors beyond adjacent swaps. A basis transformation is accepted if it strictly decreases the potential. In particular, a basis vector may be moved to an earlier position in the basis if this operation reduces $\operatorname{Pot}(B)$.

Thus, PotLLL alternates between:

\begin{itemize}
\item \textbf{Size reduction:} ensuring $|\mu_{i,j}|\le 1/2$ for $j<i$,
as in LLL;
\item \textbf{Potential-reducing insertion steps:} moving basis vectors to
earlier positions whenever this strictly decreases the global potential.
\end{itemize}

Unlike LLL, which enforces only a local Lovász condition between adjacent Gram--Schmidt lengths, PotLLL uses a global criterion that allows long-distance swaps. This often yields better-reduced bases in practice while retaining a polynomial-time complexity guarantee. Upon termination, the basis is size-reduced and locally optimal with respect to the potential function, meaning no single insertion can further decrease $\operatorname{Pot}(B)$. This typically produces bases of quality comparable to or slightly better than classical LLL, particularly in higher dimensions.

Algorithm~\ref{algo:potLLL} reproduces the PotLLL2 algorithm as presented in the original paper \cite{PotLLL2014}. Throughout this report, potLLL2 is denoted as potLLL, as only a single version is considered.

\section{The MergeLLL Construction}
MergeLLL is now formalized as a divide-and-conquer framework for lattice basis reduction, defined over a hierarchical decomposition tree inspired by merge sort.

\subsection{Recursive Decomposition}
Without loss of generality, an ${n \times n}$  basis matrix is assumed.
Let $B \in \mathbb{Z}^{n \times n}$ be a full-rank lattice basis of a lattice ${L}(B) \subset \mathbb{Z}^n$. MergeLLL recursively partitions $B$ into two sub-bases:
\[
B = B^{(1)} \cup B^{(2)},
\]
where $B^{(1)} \in \mathbb{Z}^{n_1 \times n}$ and $B^{(2)} \in \mathbb{Z}^{n_2 \times n}$ with $n_1 + n_2 = n$.

This induces a binary decomposition tree $T(B)$, where each node corresponds to a sub-basis of $B$ and the leaves correspond to sufficiently small sub-bases on which local reduction is applied.


\subsection{Local Reduction}

In BKZ, local reduction refers to improving a lattice basis by operating on short windows of vectors rather than the full basis. Given a basis $B = (b_1,\dots,b_n)$, a window $(b_k,\dots,b_{k+\beta-1})$ is projected onto the orthogonal complement of $\mathrm{span}(b_1^*,\dots,b_{k-1}^*)$ and locally reduced by (approximately) solving an SVP in the resulting sublattice. If a sufficiently short vector is found, it is inserted at position $k$, followed by size reduction.

In our application, local windows are treated as independent computational units for efficiency, allowing parallel or decoupled processing of blocks. This deviates from the standard BKZ coupling between successive blocks and is used here as a heuristic acceleration. We employ BKZ 2.0 as a black-box implementation of the underlying local reduction, using a block size equal to the sub-basis size at the leaf nodes of the hierarchical tree.
Throughout this paper, we refer to the local reduction of a sub-basis as \emph{approximate KZ reduction}, implemented using BKZ~2.0 as a black-box approximation to KZ reduction.

\subsection{Merge operation}
The merge operation can be viewed as recursively combining adjacent sub-bases. The leaf sub-bases are first approximately reduced using BKZ, while every merge between adjacent sub-bases is performed using PotLLL. Suppose the following two blocks are to be merged
\[
B_1 = (b_1,\dots,b_{n_1}), \quad B_2 = (b_{n_1+1},\dots,b_{n_2}),
\]
where both $B_1$ and $B_2$ are independently reduced according to their position in the reduction tree: leaf sub-bases are reduced using approximate KZ-reduction, while intermediate sub-bases are meregd and reduced using PotLLL. The only modification needed in the original potLLL algorithm for these two inputs is to start from \(l = n_1 + 1\), and to apply the algorithm to the concatenated basis
\[
B = B_1 \cup B_2,
\]
where \(B\) is obtained by concatenating \(B_2\) after \(B_1\). In addition, the initial preprocessing step of PotLLL using classical LLL reduction is omitted.
The merge operation is outline in Algorithm \ref{Algo:Merge}.

Although each block \(B_1\) and \(B_2\) satisfies the potLLL (or BKZ) invariants internally, this does not imply that no further internal modifications will occur during the final merge. However, the hierarchical structure is still useful: the previous merging steps significantly reduce the number of violations and ensure that the final merge begins from a highly structured state. Consequently, while the final merge is not strictly confined to boundary corrections, it typically performs substantially fewer swap operations than running potLLL from scratch.

It can be argued that the previous merging steps may, in the worst case, be wasteful, in the sense that their overall effect can be comparable to restarting the reduction from scratch on the full basis. In particular, although locally reduced blocks are produced by these intermediate merges, global optimality with respect to the final merged basis is not necessarily preserved, and hence the total work required by the final reduction stage may not be reduced. However, empirically, these steps are observed to improve performance in practice.

From a parallelization perspective, the earlier merges can be executed independently on disjoint sub-bases, and the algorithm becomes sequential only at the final merge stage. 

\begin{algorithm}[H]
\caption{\textsc{MergeReducedBases} (PotLLL Insertion of a Basis $B_2$ into $B_1$)}
\begin{algorithmic}[1]

\Require PotLLL-reduced basis $B_1 = (b_1,\dots,b_{n_1})$, and
$B_2 = (b_{n_1+1},\dots,b_{n_2})$, parameter $\delta \in (1/4,1)$
\Ensure $\delta$-PotLLL reduced basis B
\State Set $B \gets B_1 || B_2$ \Comment{concatinate $B_1$ and $B_2$}

\State $l \gets n_1+1$
\State $n \gets n_1+n_2$

\While{$\ell \le n$}
    \State Size-reduce $b_\ell$ by $b_1,\ldots,b_{\ell-1}$
    \State Update $\|b_\ell^*\|^2$ and $\mu_{\ell,j}$ for $1 \le j < \ell$

    \State $P \gets 1$
    \State $P_{\min} \gets 1$
    \State $k \gets 1$

    \For{$j=\ell-1$ downto $1$}
        \State
        $P \gets P \cdot
        \dfrac{
            \|b_\ell^*\|^2 +
            \sum_{i=j}^{\ell-1}
            \mu_{\ell,i}^2 \|b_i^*\|^2
        }{
            \|b_j^*\|^2
        }$
        \If{$P < P_{\min}$}
            \State $k \gets j$
            \State $P_{\min} \gets P$
        \EndIf
    \EndFor

    \If{$\delta > P_{\min}$}
        \State $B \gets \sigma_{k,\ell} B$
        \State Update $\|b_k^*\|^2$ and $\mu_{k,j}$ for $1 \le j < k$
        \State $\ell \gets k$
    \Else
        \State $\ell \gets \ell + 1$
    \EndIf
\EndWhile

\State \Return $B$

\end{algorithmic} \label{Algo:Merge}
\end{algorithm}

\subsection{Correctness of Merging Operation}

Let \(B_1 = (b_1,\dots,b_{n_1})\) and \(B_2 = (b_{n_1+1},\dots,b_{n_2})\) be two bases that are independently potLLL-reduced (or BKZ-reduced), and let \(B = B_1 \cup B_2\) denote their concatenation.

Since both \(B_1\) and \(B_2\) are already locally-reduced, all local reduction conditions (size-reduction and potential-decrease conditions) are satisfied within each block. 

By starting potLLL from \(k = n_1 + 1\) implies that:
\begin{itemize}
\item the internal structure of \(B_1\) is not immediately revisited, since all pairs \((b_i,b_{i+1})\) for \(i < n_1\) already satisfy the potLLL conditions at initialization;
\item only interactions introduced by the concatenation are directly acted upon at the start, although subsequent deep insertions may still propagate updates into earlier indices.
\end{itemize}

Each swap or size-reduction step performed by potLLL strictly decreases the global potential function, which is bounded from below. Therefore, the algorithm terminates after finitely many steps. Upon termination, all adjacent pairs in \(B\) satisfy the potLLL conditions, including those near the boundary. Hence, the output basis is potLLL-reduced and spans the same lattice as \(B_1 \cup B_2\), establishing correctness of the merging operation.

\subsection{The MergeLLL Algorithm}
The MergeLLL algorithm is outlined in Algorithm~\ref{Algo:Merge}. It describes a divide-and-conquer procedure in which the basis is recursively split into smaller blocks, reduced independently, and then iteratively merged. At each merging step, adjacent blocks are combined and lattice reduction is applied to the concatenated basis. This strategy is designed to progressively improve the structure of the basis.

\begin{algorithm}[H]
\caption{MergeLLL (with Approximate KZ Initialization and merge using potLLL)}
\label{algo:MergeLLL}
\begin{algorithmic}[1]
\Require Basis $B=(b_1,\ldots,b_n)$, block size $\beta$, parameter $\delta\in(1/4,1)$
\Ensure $\delta$-PotLLL reduced basis $B$

    \State $n \gets |B|$

    \Comment{Initial local reductions}
    \For{$left \gets 1$ \textbf{to} $n$ \textbf{step} $\beta$}
        \State $right \gets \min(left+\beta-1,n)$
        \State $B[left\ldots right] \gets \textsc{ApproxKZReduce}(B[left\ldots right])$
    \EndFor

    \State $size \gets \beta$

    \Comment{Bottom-up merge phase}
    \While{$size < n$}
        \For{$left \gets 1$ \textbf{to} $n$ \textbf{step} $2\cdot size$}
            \State $mid \gets \min(left+size-1,n)$
            \State $right \gets \min(left+2\cdot size-1,n)$

            \If{$mid < right$}
                \State $B[left\ldots right] \gets \textsc{MergeReducedBases}(B,left,mid,right)$
            \EndIf
        \EndFor
        \State $size \gets 2\cdot size$
    \EndWhile
\end{algorithmic}
\end{algorithm}

\section{Experimental Evaluation}
MergeLLL is evaluated as a framework for lattice basis reduction, with emphasis placed on its effect on basis quality and computational behavior. The experiments are conducted on integer lattices of moderate to high dimensions, including instances derived from subset-sum constructions and NTRU-style structured lattices.


\subsection{Experimental Setup}
MergeLLL is implemented as described in Algorithm~\ref{algo:MergeLLL} and in a parallelized form. Within the same framework, the potLLL~\ref{algo:potLLL} is also implemented following its original specification. The BKZ-reduction is performed using the BKZ 2.0 algorithm, as implemented in the C++ \texttt{fplll} library.

All experiments are implemented in C++ and run on the PDC (KTH) high-performance computing cluster. Computations are carried out on the \texttt{main} partition under the project account \texttt{naiss2026-4-509}. Each run uses a single compute node with 64 CPU cores. The source code is available at \url{https://github.com/niharikag/MergeLLL}

In all experiments, the reduction parameter is set to $\delta = 0.99$.
All figures in this paper are generated using the Python Matplotlib library.

\subsection{Evaluation Metrics}
In this work, the root Hermite factor (RHF) is considered as measures for evaluating a lattice basis. These quantities capture both the lengths of the basis vectors and the degree to which the basis approaches orthogonality. They are particularly useful for analyzing the performance of lattice reduction algorithms.

\begin{itemize}

\item \textbf{Root Hermite's Factor (RHF):} The Hermite factor of a basis $B$ is defined as
\[
\mathrm{HF}(B)
=
\frac{\|b_1\|}
{\det(L)^{1/n}},
\]
where $b_1$ is the shortest vector of the basis $B$.  The n-th root of the HF are reported in all experiments. This quantity is always greater than or equal to 1, and smaller values indicate better reduced bases, with values closer to 1 corresponding to higher-quality reductions.

\item \textbf{CPU Time:} The reported CPU time corresponds to the total time required to execute the full lattice reduction procedure, including potLLL and, in the case of MergeLLL, all splitting and hierarchical merging operations. 
\end{itemize}

\subsection{Test Instances}
PotLLL and MergeLLL are compared under identical experimental conditions, while BKZ 2.0 is executed using the optimized implementation provided by the C++ fplll library. The evaluation is conducted on two complementary families of lattices with varying dimensions to assess scalability and numerical stability.

\begin{itemize}
    \item \textbf{Subset-sum lattices:} combinatorial and largely unstructured instances constructed using the standard subset-sum embedding as in the BKZ benchmarking framework\cite{BKZ2011}. The lattice dimension ranges from \(21\) to \(401\) in increments of \(20\). The subset-sum instances use bit-length parameters \(20\).
    
\vspace{5pt}

    Let $a=(a_1,\dots,a_n)\in\mathbb{Z}^n$ and $S=\sum_{i=1}^n a_i x_i$ for $x_i\in\{0,1\}$. A  basis $B\in\mathbb{Z}^{(n+1)\times(n+2)}$ is constructed with rows $b_1,\dots,b_{n+1}$.
    
    For $i=1,\dots,n$,
    \[
    b_i = (0,\dots,0,2,0,\dots,0, n a_i, 0),
    \]
    where $2$ is in position $i$.
    
    The additional vector is
    \[
    b_{n+1} = (1,1,\dots,1, nS, 1).
    \]
    
    The vector $b_{n+1}$ enforces the subset-sum constraint within the lattice embedding.

    \item \textbf{NTRU-style lattices:} structured lattices arising from ring-based NTRU constructions, exhibiting strong algebraic dependencies induced by convolution structure in polynomial rings. Specifically, the lattice basis has the form
    \[
    \begin{pmatrix}
    q I & 0 \\
    H & I
    \end{pmatrix},
    \]
    where $H$ is a circulant matrix generated from a random polynomial $h$. In the experiments, $q = 2^{31} - 1$ is used. The dimension is varied from 20 to 100 in increments of 20.
\end{itemize}


\subsection{Experiment 1}
\begin{figure}[H]
    \centering
    \includegraphics[width=0.7\linewidth]{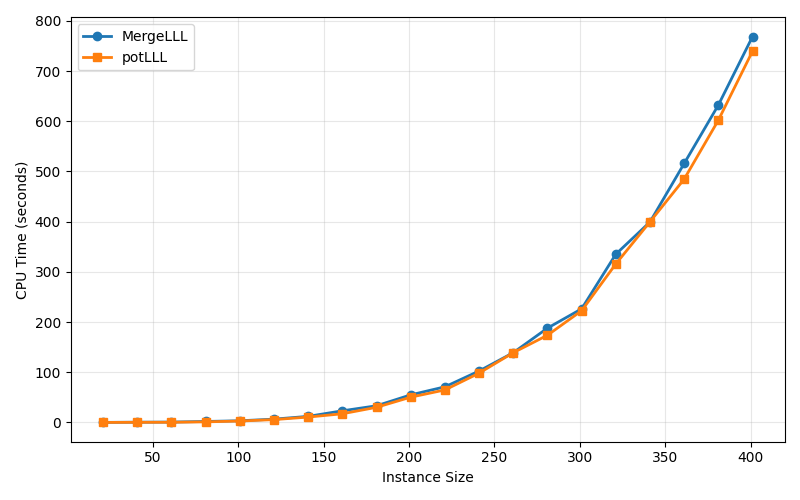}
    \caption{CPU time comparison on Subset Sum instances with weights of bit-length 20}
    \label{fig:subsetsum_20bits}
\end{figure}

In this experiment, MergeLLL is evaluated on subset sum lattice bases generated using 20-bit integer weights and compared with the classical LLL algorithm. The instance size is varied from 20 to 400, and CPU time and Root Hermite Factor (RHF) are measured. The leaf node size and block size are both set to 10. The results are reported in Table~\ref{table:subsetsum_20bits} and Figure~\ref{fig:subsetsum_20bits}.

\begin{table}[h]
\centering
\begin{tabular}{c c c c}
\hline
Dim & LLL\_RHF & MergeBKZ\_RHF & BKZ\_RHF \\
\hline
21  & 1.00091 & 1.00091 & 1.00091 \\
41  & 1.00645 & 1.00645 & 1.00919 \\
61  & 1.00835 & 1.00835 & 1.00835 \\
81  & 1.00388 & 1.00388 & 1.00705 \\
101 & 1.00501 & 1.00501 & 1.00501 \\
121 & 1.00442 & 1.00442 & 1.00535 \\
141 & 1.00474 & 1.00394 & 1.00474 \\
161 & 1.00355 & 1.00355 & 1.00425 \\
181 & 1.00323 & 1.00323 & 1.00323 \\
201 & 1.00352 & 1.00296 & 1.00352 \\
221 & 1.00273 & 1.00273 & 1.00273 \\
241 & 1.00253 & 1.00299 & 1.00253 \\
261 & 1.00235 & 1.00235 & 1.00278 \\
281 & 1.00221 & 1.00221 & 1.00261 \\
301 & 1.00208 & 1.00208 & 1.00208 \\
321 & 1.00196 & 1.00196 & 1.00196 \\
341 & 1.00143 & 1.00143 & 1.00185 \\
361 & 1.00176 & 1.00176 & 1.00176 \\
381 & 1.00168 & 1.00168 & 1.00168 \\
401 & 1.00124 & 1.00124 & 1.00160 \\
\hline
\end{tabular}
\caption{RHF values for different dimensions and reduction methods} \label{table:subsetsum_20bits}
\end{table}

\subsection{Experiment 2}
\begin{figure}[ht]
    \centering
    \includegraphics[width=0.7\linewidth]{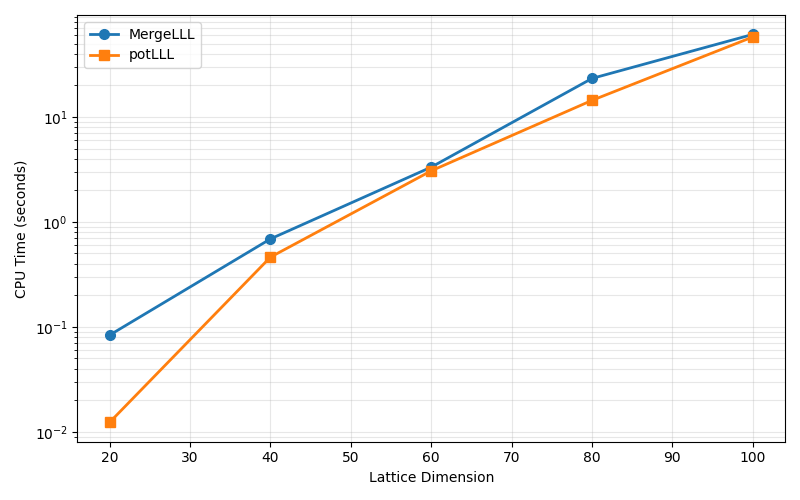}
    \caption{CPU time comparison on NTRU-style lattice instances} \label{fig:NTRU}
\end{figure}

The experimental setup is identical to that of Experiment 1, except that NTRU-style lattice bases are used instead of subset sum bases. The lattice dimension is varied from 20 to 100. The leaf node size and block size are both set to 20. 
The results are reported in Table \ref{table:NTRU} and Figure \ref{fig:NTRU}.

\begin{table}[ht]
\centering
\begin{tabular}{c c c c}
\hline
Dim & LLL\_RHF & MergeLLL\_RHF & BKZ\_RHF \\
\hline
20  & 1.01739 & 1.01739 & 1.01739 \\
40  & 1.01402 & 1.01402 & 1.01402 \\
60  & 1.01440 & 1.01325 & 1.01368 \\
80  & 1.01531 & 1.01355 & 1.01371 \\
100 & 1.01400 & 1.01415 & 1.01395 \\
\hline
\end{tabular}
\caption{RHF values for LLL, MergeBKZ, and BKZ} \label{table:NTRU}
\end{table}

\subsection{Discussion}
The experimental results support the main design principle of MergeLLL: lattice reduction can be effectively decomposed into local reduction followed by controlled global recombination. By isolating most of the reduction effort into independent sub-bases, the framework enables substantial parallelism while maintaining strong overall basis quality.

As an initial evaluation, the results indicate that the RHF performance metric is comparable to PotLLL and outperforms BKZ-10 across all experiments. The CPU time is also comparable to potLLL.

For Subset Sum instances with 20-bit bases and dimensions up to 400, MergeLLL shows performance comparable to PotLLL in terms of both CPU time and root Hermite factor (RHF). Compared to BKZ (using an optimized implementation), MergeLLL achieves better RHF values; however, we do not report a direct runtime comparison with BKZ due to its reliance on an external optimized library implementation.

For NTRU-style bases, a similar behavior to Subset Sum instances is observed. The reduction quality, in terms of root Hermite factor, is comparable between the methods, with MergeLLL maintaining stable behavior and producing competitive reduced bases.

Overall, the results demonstrate that MergeLLL provides a scalable approach for structured lattice families such as Subset Sum and NTRU-style instances, making it suitable for higher-dimensional settings where computational costs become significant.

\section{Conclusion and Future Work}\label{sec:conclusion}
In this work, we propose and evaluate the MergeLLL framework for lattice basis reduction, and compare it against the PotLLL algorithm on Subset Sum and NTRU-style lattice instances. It exhibits performance comparable to PotLLL in terms of both CPU time and reduction quality, while achieving better reduction quality than BKZ under the tested settings.

Despite the independent reductions, the quality of the resulting bases remains comparable, as indicated by similar root Hermite factors. This suggests that the proposed hierarchical merge strategy provides an effective trade-off between computational efficiency and reduction quality.

Overall, the experimental results indicate that MergeLLL is a practical and scalable alternative for structured lattice families, especially in regimes where standard LLL becomes infeasible due to computational cost.

It is believed that incorporating a true KZ reduction at the leaf level may further improve performance and potentially enable the method to surpass PotLLL in terms of reduction quality.

\subsection{Future Work} 
The current implementation is written in C++, where the basis matrix entries are represented using the GMP library's arbitrary-precision integer type (\texttt{mpz\_class}), while the Gram--Schmidt orthogonalization and the corresponding coefficients are computed using the GMP arbitrary-precision floating-point type (\texttt{mpf\_class}).

\vspace{5pt}

\noindent Future directions include:
\begin{itemize}
    \item Extending the implementation to use arbitrary-precision arithmetic via MPFR (Multiple Precision Floating-Point Reliable) library\cite{mpfr2007} to improve numerical stability for large instances.
    \item Evaluating the algorithm on SVP Challenge lattice bases\cite{svpchallenge} to test performance on standard benchmark instances.
\item  Incorporating original KZ/BKZ reduction at the leaf nodes with varying block sizes to study quality–performance trade-offs.
\end{itemize}

\backmatter

\section{Supplementary information}
Not Applicable.

\section{Acknowledgements}
The support of the PDC Center for High Performance Computing at KTH Royal Institute of Technology is gratefully acknowledged. The computational resources used in this work are provided under the project \texttt{naiss2026-4-509}.

\section*{Declarations}
Not Applicable.

\bibliography{sn-bibliography}

\end{document}